\documentclass[twocolumn,showpacs,preprintnumbers,amsmath,amssymb]{revtex4}

\usepackage{graphicx}
\usepackage{dcolumn}
\usepackage{bm}

\begin{document}

\preprint{APS/123-QED}

\title{Anisotropic scaling of magnetohydrodynamic turbulence}

\author{Timothy S. Horbury}
 \email{t.horbury@imperial.ac.uk}
 \homepage{http://www.imperial.ac.uk/people/t.horbury}
\affiliation{%
The Blackett Laboratory
Imperial College London\\
SW7 2AZ
U.K.
}%

\author{Miriam Forman}
\affiliation{Stony Brook University, Stony Brook, N. Y. 11794, U.S.A.
}%

\author{Sean Oughton}
\affiliation{Department of Mathematics, University of Waikato, Hamilton, New Zealand
}%

\date{\today}

\begin{abstract}
We present a quantitative estimate of the anisotropic power and scaling of magnetic field fluctuations in inertial range magnetohydrodynamic turbulence, using a novel wavelet technique applied to spacecraft measurements in the solar wind. We show for the first time that, when the local magnetic field direction is parallel to the flow, the spacecraft-frame spectrum has a spectral index near 2. This can be interpreted as the signature of a population of fluctuations in field-parallel wavenumbers with a $k_{\parallel}^{-2}$ spectrum but is also consistent with the presence of a ``critical balance" style turbulent cascade. We also find, in common with previous studies, that most of the power is contained in wavevectors at large angles to the local magnetic field and that this component of the turbulence has a spectral index of $5/3$. 
\end{abstract}

\maketitle

Magnetised plasmas fill most of the Universe and in many regions, turbulence plays important roles in the transport of energy and momentum and the acceleration and scattering of charged particles. Many aspects of plasma turbulence remain poorly understood, however. Here we present results on one of these, the anisotropy of the energy spectrum of magnetohydrodynamic (MHD) turbulence with respect to the magnetic field.

In classical hydrodynamics, velocity fluctuations $\delta u_{k}$ with a wavenumber $k$ decay and transfer energy to smaller scales on the shear timescale, $\tau_{S} \approx 1/ (k \delta u_{k})$. Within the steady inertial range, far from the energy input (``outer") and dissipation scales, this leads to the dimensional result $(\delta u_{k})^{3} \propto \epsilon/k$, where $\epsilon$ is the energy dissipation rate per unit mass. This gives the familiar Kolmogorov energy spectrum $P(k) \propto k^{-5/3}$, widely observed in hydrodynamic turbulence. In a plasma, fluctuations can also propagate, as Alfv\'en waves parallel to the magnetic field, and this leads to the Alfv\'en timescale, $\tau_{A} \approx 1/(k_{\parallel} V_{A})$, being dynamically important. Here $k_{\parallel}$ is the component of the wavevector of the fluctuation parallel to the local magnetic field and $V_{A}$ the Alfv\'en speed. If $\tau_{A} \ll \tau_{S}$ and assuming isotropy with respect to the local field, this leads to Iroshnikov-Kraichnan turbulence \citep{Iroshnikov63,Kraichnan65} where $(\delta u_{k})^{4} \propto \epsilon V_{A} / k$ and $P(k) \propto k^{-3/2}$ \citep[e.g.][]{Biskamp-turb,SchekochihinEA07-ppcf}.

However, measurements in both space plasmas and terrestrial plasma devices have shown that turbulent fluctuations are not isotropic. They typically have much longer correlation lengths along the field than across it \citep{RobinsonRusbridge71,ZwebenEA79,MattEA90,CarboneEA95-Rij,DassoEA05,OsmanHorbury07} and the spectral index for the magnetic energy is nearer 5/3 than 3/2 \citep{PodestaEA07-slopes, BrunoCarbone05}. When there is an energetically significant large-scale magnetic field, anisotropic models of MHD turbulence are required \citep{Higdon84,Mont82-strauss,ChoEA02-strongB0,GoldreichSridhar95,Boldyrev06,MasonEA08}. For example, in the ``critical balance" framework \citep{GoldreichSridhar95}, turbulent energy evolves towards wavevectors  where the shear and Alfv\'en timescales are balanced and most power resides in wavevectors where $\tau_{S} \le \tau_{A}$, i.e. $k_{\parallel} \le k_{\perp}^{2/3} \epsilon^{1/3} V_{A}^{-1}$.

The solar wind is a unique environment in which to study space plasma turbulence: it is relatively accessible and can be directly measured in exquisite detail using spacecraft instruments \citep[e.g.][]{TuMarsch95,MattEA95-unquiet,BrunoCarbone05,HorburyEA05}. The solar wind flows radially away from the Sun at a velocity $\bm{V}$ of several hundred km\,s$^{-1}$, much faster than spacecraft motions (a few km\,s$^{-1}$) or the plasma wave speeds (tens of km\,s$^{-1}$). As a result, in the plasma frame spacecraft measure along a radial line. Using Taylor's hypothesis \citep{Taylor38}, one can relate the spacecraft frame energy
spectrum $ \mathcal{P}(f) $ to the wavevector spectrum $P(\bm{k})$ \citep{FredricksCoroniti76}:       
\begin{eqnarray}
        \label{RedSpecEqn}
 \mathcal{P} (f)
   =
 \int d^{3} \bm{k} \,
        P(\bm{k})
        \, \delta (2 \pi f - \bm{k} \cdot \bm{V}) .
\end{eqnarray}
Anisotropies in $P(\bm{k})$ with respect to the magnetic field can be analysed by measuring how $\mathcal{P}(f)$ varies with the angle
of the magnetic field to the flow,
      $ \theta_{B} $.
      
The exact form of this anisotropy is unknown, but one can make approximations motivated by theory and compare predictions with observations. One simple approximation is to assume that $P(\bm{k})=0$ except for wavevectors exactly parallel (so-called ``slab") or perpendicular (``2D") to the local magnetic field \citep{BieberEA96}. The corresponding frequency spectrum can be deduced from Eq. \ref{RedSpecEqn}:
\begin{eqnarray}
 \mathcal{P} (f ; \theta_{B})
   = 
 C_{slab} f^{-\gamma_{slab}}
          \left| \cos \theta_{B} \right|^{\gamma_{slab}-1} \nonumber \\ 
          +  C_{2D} f^{-\gamma_{2D}}
          \left| \sin \theta_{B} \right|^{\gamma_{2D}-1},
\end{eqnarray}
where $C_{slab}$ and $C_{2D}$ are constants and $\gamma_{slab}$ and $\gamma_{2D}$ are the spectral indexes of these components. $\mathcal{P}$ is thus insensitive to the slab component when the field is perpendicular to the flow and insensitive to 2D when the field is parallel \citep{BieberEA96}. Crucially, one can determine the scaling of both components by measuring the spectral index $\alpha$ of $\mathcal{P} \approx f^{-\alpha}$ separately for $\theta_{B}=0^{\circ}$ and $\theta_{B}=90^{\circ}$.

In the case of critical balance \citep{GoldreichSridhar95}, the 3D power spectrum takes the form
\begin{eqnarray}
\label{GSspec}
 P(\bm{k}) \propto k_{\perp}^{-10/3} g \left( \frac{ V_{A} k_{\parallel}}{\epsilon^{1/3} k_{\perp}^{2/3}} \right) .
\end{eqnarray}
$\mathcal{P}(f ; \theta_{B})$ then depends in a complicated way on the unspecified function $g(y)$, but one can show that Eq. \ref{GSspec} implies $\mathcal{P}(f ; \theta_{B}=90^{\circ}) \propto f^{-5/3}$ and $\mathcal{P}(f ; \theta_{B}=0^{\circ}) \propto f^{-2}$, with the latter also smaller in magnitude at a given $f$. This result is independent of the precise form of $g(y)$. In the case of critical balance we would therefore expect an anisotropy in both the power levels and the spectral index of the spectrum.

Here, we use 30 days (1995, days 100-130) of 1 second resolution measurements of magnetic field fluctuations \citep{BaloghEA99} by the Ulysses spacecraft. During this time, Ulysses was within the steady high speed ($V \sim 750\,\textnormal{km\,s}^{-1}$) solar wind from the Sun's Northern polar
coronal hole at 1.4\,AU from the Sun \citep{McComasEA00}. Fluctuations within the solar wind inertial range, corresponding to
spacecraft time scales of seconds to minutes
        \citep{HorburyBalogh01},
are superimposed on large amplitude
        ($| \delta \bm{B} | / | B | \sim 1$)
Alfv\'en waves on time scales of hours
        \citep{SmithEA95d}
which result
in large variations in $\theta_{B}$.  The minimum variance direction
of the inertial range fluctuations follows the local magnetic field
direction very closely
        \citep{HorburyEA95c},
indicating that the local field orders the behaviour of the fluctuations. We can therefore study the anisotropies of the turbulence by measuring how the spacecraft frame spectrum of magnetic fluctuations varies with $\theta_{B}$. We perform this analysis using a new wavelet method, which is sensitive to the constantly changing local magnetic field direction. 

Wavelets have been used extensively to study physical time series \citep[e.g.][]{PercivalWalden,Farge92,TorrenceCompo98,AlexandrovaEA06}. The Morlet wavelet is
relatively well-localised in frequency, being rather wave-like
  \citep{PercivalWalden}:
\begin{eqnarray}
  \psi (x)
    & = &
  \pi^{-1/4}  e^{i \omega_0 x} e^{-x^{2}/2}.
                                        \label{eq:motherMorlet}
\end{eqnarray}
With $\omega_0 = 6$, it is possible to construct a nearly orthonormal set of wavelets. For each magnetic field component $i$, we have a
time series
        $ B_{i} (t_k)$,
where   $ t_k = t_{0} + k \delta t $
and
        $ \delta t = 1$\,s. The discrete wavelet transform $w_i (t_j, f_l)$ of such a time series, at a time $t_j$ and frequency $f_l$ is given by
\begin{eqnarray}
                \label{DWTEqn}
  w_i (t_j, f_l)
  =
    \sum_{k=0}^{N-1}
          B_{i}(t_k)
        \,
          \psi \left( \frac{t_k - t_j}{s_l} \right)
  .
\end{eqnarray}
The time scale or dilation parameter $s_{l}$ is directly related to the peak frequency response $f_{l}$ of the wavelet. For $\omega_0=6$, $s_l \doteq 1.03 1/f_l$ \citep[e.g.][]{TorrenceCompo98}. In practice, the wavelet transform is more efficiently calculated using Fourier transforms rather than directly  in the time domain. We calculate the wavelet coefficients at ten logarithmically spaced frequencies $f_{l}=f_{0} \cdot (8/5)^{-l}$, where $l=0, 1, \ldots, 9$: $f_{0}=0.25$~Hz and $f_{9}=3.6$~mHz.

The wavelet coefficients $w_{i}$ can be used to calculate the power in a time series: at a time $t_j$ and frequency $f_l$ in component $i$ the power is proportional to
\begin{eqnarray}
 \mathcal{P}_{ii} (t_j, f_l) \propto f_{l} \left| w_{i}(t_{j},f_{l}) \right| ^{2} .
 \end{eqnarray}
Here we analyse the trace, $\mathcal{P}=\Sigma \mathcal{P}_{ii}$. 

Fluctuations at a given
scale 
are sensitive to the local magnetic field -- but the definition of ``local" varies with the spatial scale of the fluctuations of interest. In general, one would expect fluctuations with a given wavelength to be sensitive to the magnetic field on approximately this scale and above. 

In order to measure the scale-dependent local magnetic field direction, we calculate the amplitude envelope of the Morlet wavelet, $| \psi (t_j,f_l)|^2$ -- this is a Gaussian centered on time $t_j$ with a width of $1.67 s_l$ -- and calculate the sum over the data set of the product of this envelope with the magnetic field time series, for each field component $i$:
\begin{eqnarray}
 b_{i} (t_{j}, s_{l}) = \sum_{k=0}^{N-1}
          B_{i}(t_k)
       \,
          \left| \psi \left( \frac{t_k - t_j}{s_l} \right) \right| ^2 .
\end{eqnarray}
For each  $s_{l}$, this results in a time series of vectors $\bm{b}(t_{j},s_{l})$ which point in the direction of the local magnetic field at the time $t_j$, associated with a time scale $s_{l}$ (or equivalently $f_{l}$). This provides a frequency- and time-localised mean field direction, and hence $\theta_{B}$ and the azimuthal angle $\phi_{B}$, for every wavelet coefficient. 

We next construct a set of 404 bins, each of which subtends approximately the
same solid angle, and which together cover all directions.  For each bin,
for a given frequency $f_{l}$ and field component $i$, we select all wavelet
coefficients that have average magnetic field angles
($\theta_{B}$,$\phi_{B}$) within the bin.  The mean of these
coefficients is then the average power in component $i$ at frequency $f_{l}$ when the field points in that
direction. 

This process results in a scale-sensitive estimate of the
magnetic field power spectrum as a function of the magnetic field
angle relative to the solar wind flow (sampling) direction - that is,
we estimate
        $ \mathcal{P}(f; \theta_{B}, \phi_{B}) $.
Many bins contain thousands of
measurements; in order to ensure
reliable statistics, any angle bin with fewer than 40 contributing
power levels is rejected. If the fluctuations are axisymmetric around $\bm{B}$, $ \mathcal{P}$ should be independent of  $\phi_{B}$.
Our measurements indicate that this is indeed the case and we therefore consider values averaged over all $\phi_{B}$.

\begin{figure}
\includegraphics{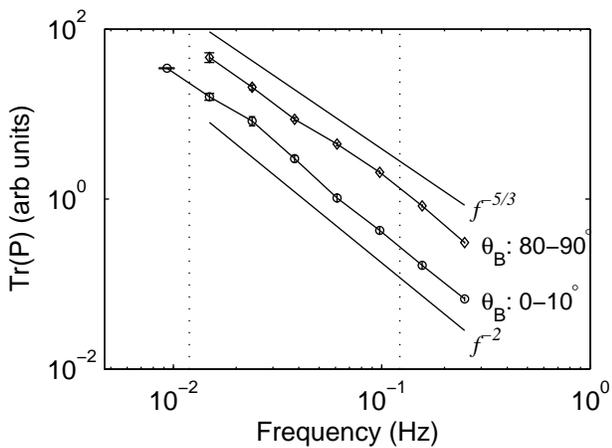}
\caption{\label{fig:powspect} Magnetic power spectra at two different angle ranges of the local
     magnetic field to the flow:
        $0$--$10^{\circ}$ (circles) and $80$--$90^{\circ}$ (diamonds).
     Note the reduced power levels and steeper slope
     associated with the smaller angle.
     Guide lines with slopes of 5/3 and 2 are shown above and below
     the data. Spectral indices in Fig.~\ref{fig:scaling} are calculated over the scales between the dotted vertical lines. 
}
\end{figure}

Using this wavelet method, we can estimate the
spacecraft-frame power spectrum at a range of frequencies $f$
and magnetic field/flow angles $\theta_{B}$.
        Fig.~\ref{fig:powspect}
shows typical power levels for two ranges of
        $\theta_{B}$:
        $0^\circ$--$10^\circ$
and
        $80^\circ$--$90^\circ $.
Although both are well described by power laws in $f$ over the range of
frequencies considered here, it is apparent that the
power levels for $\theta_{B} =  0^\circ$--$10^\circ$
are lower than those
for
        $80^\circ$--$90^\circ $, in agreement with several previous studies \citep{BieberEA96,Smith-sw10,LeamonEA98a} and consistent with the expectation that most power in the fluctuations is in wavevectors at large angles to the magnetic field \citep{MattEA90}. In addition, the spectrum for
        $\theta_{B} = 0^\circ$--$10^\circ$
is steeper than that for larger angles. 

\begin{figure}
   \includegraphics{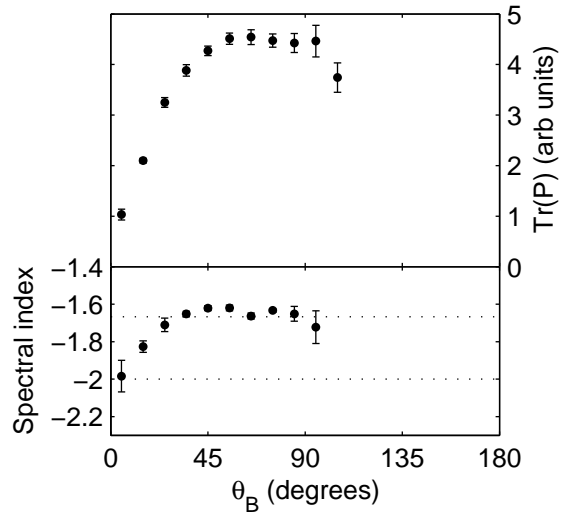}
  \caption{\label{fig:scaling}%
    Top panel: Trace of power in the magnetic
    field as a function of the angle between the local magnetic field
    and the sampling direction at a spacecraft frequency of 61\,mHz.
    The larger scatter for $\theta_{B} > 90^{\circ}$ is the
    result of fewer data points at these angles. Bottom panel: spectral
    index of the trace, fitted over spacecraft frequencies from
        15--98\,mHz.
  }
\end{figure}

The variations in the spectrum power level and spectral index with $\theta_{B}$ are more easily seen in Fig.~\ref{fig:scaling}, where it is apparent that there is a smooth variation in power with field/flow angle. Note that occasional folds in the magnetic field past $\theta_{B}=90^{\circ}$ (due to the presence of large amplitude Alfv\'en waves \citep{BaloghEA99}) mean that it is possible to measure variations for $\theta_{B} > 90^{\circ}$, although not all angles can be measured.

The most important result in this paper is shown in the bottom panel of Fig.~\ref{fig:scaling}, where it is clear that there is a systematic variation in spectral index with $\theta_{B}$. For most angles, $\alpha \approx 5/3$, in accord with most previous solar wind measurements \citep[e.g.][]{BrunoCarbone05}. However, when $\theta_{B} \to 0^{\circ}$, $\alpha$ changes in a smooth manner towards a value of around $2$. A spectral index of 2 at small field/flow angles has not previously been reported. It is strong evidence of anisotropic energy transfer in the MHD cascade and was predicted for any ``critical balance" type of cascade \citep{ChoEA02-strongB0,Boldyrev05}.
A spectral index of 2 at $\theta_{B}=0^{\circ}$ and 5/3 at $\theta_{B}=90^{\circ}$ is consistent with the presence of a critical balance cascade. However, it is surprisingly difficult to distinguish between the critical balance and slab/2D approximations from these results and indeed they are also broadly consistent with a dominant population of 2D fluctuations with $\gamma_{2D}=5/3$ and a smaller slab population with $\gamma_{slab}=2$. The small range of scales over which we measure the fluctuations does not rule out other scalings (e.g.\ exponential) for a possible low amplitude slab component. 

It is perhaps surprising that the $f^{-2}$ scaling has not previously been observed in the solar wind and indeed we are aware of ongoing work by others which does not show such variation. However, the constantly changing background field direction means that only by using very short averaging periods can we avoid ``smearing out" the very low power fluctuations observed when $\theta_{B} \approx 0$, which is the only time that we see the $f^{-2}$ scaling. Recent multi-spacecraft studies (which are not susceptible to these effects) have not revealed this scaling in the solar wind \citep{NaritaEA07-riverside}, although anisotropic scaling has been observed in the magnetosheath \citep{SahraouiEA06}.

Note that the variation of $\mathcal{P}(\theta_{B})$ in Fig.~\ref{fig:scaling} is not symmetric around $\theta_{B}=90^{\circ}$, which is not possible under our assumptions of homogeneity and Taylor's hypothesis, and therefore one or both of these assumptions must to some extent be violated. The lack of symmetry in Fig.~\ref{fig:scaling} may be due to kinetic effects at the small scale edge of the inertial range, although Fig.~\ref{fig:powspect} suggests that the steeper spectrum for small $\theta_{B}$ extends over a wide range of frequencies. The dynamical effects which cause local changes in the field direction, including large-scale Alfv\'en waves and microstreams, might cause systematic changes in power levels with $\theta_{B}$. However, while they might change $\mathcal{P}(\theta_{B})$, they would be unlikely to change the $\alpha$: we consider the measurement of a steeper spectrum near $\theta_{B}=0^{\circ}$ to be a robust result.

We are only measuring the magnetic energy spectrum here, and not that of the plasma velocity: telemetry limitations make it impossible to study the velocity fluctuations on these timescales using Ulysses data. However, other observations \citep{PodestaEA07-slopes} indicate that the solar wind magnetic and kinetic energy spectra often have distinct slopes, typically 5/3 and 3/2 respectively. Many simulation studies also find distinct slopes, as does a recent closure model \citep{MullerGrappin05}, but most theoretical models give forms for the total (kinetic plus magnetic) energy spectrum, making it difficult to compare them directly with our results. Nonetheless, our results are in accord with simulations where the amplitude of the background field is approximately the same as the rms $\bm{B}$-- as occurs in high-latitude solar wind \citep{ChoVishniac00-aniso} -- and which yield kinetic and magnetic spectra with slopes of $\approx 5/3$. Our results are not consistent with spectral indexes of $3/2$ associated with strong background field simulations and models \citep[e.g.,][]{MaronGoldreich01,MullerGrappin05,Boldyrev06}.

The anisotropy reported here may also be influenced by the anisotropic ``driving'' of the turbulence at large scales by large-amplitude, predominantly anti-Sunward-propagating, Alfv\'en waves.  Such driving is in contrast to the probably isotropic and possibly weak-amplitude injection occurring in many astrophysical plasmas, and may limit the applicability of our findings to other plasma regimes.

The range of $\theta_{B}$ over which the spectral index deviates from $5/3$ is rather larger than we would expect on the basis of the variation of $\mathcal{P}$ with $\theta_{B}$, under either a critical balance or slab/2D cascade framework. Currently we do not have a good explanation for this discrepancy but hope to address it in a later paper.

Finally, we note that the wavelet method can be extended beyond just the trace of the spectrum as used here. We can also measure the individual elements of the power spectral tensor, revealing information about the field-parallel variance (diagonal elements), helicity (off-diagonal elements) and even intermittency (using higher-order moments). We intend to present our analyses of these measurements in the near future.

Ulysses analysis at Imperial College London is supported by the UK STFC. 
The authors are grateful to ISSI,
Bern, for their support of this work.  T. Horbury acknowledges helpful
conversations with A. Schekochihin, K. Osman and C. Chen.  


\begin{thebibliography}{41}
\expandafter\ifx\csname natexlab\endcsname\relax\def\natexlab#1{#1}\fi
\expandafter\ifx\csname bibnamefont\endcsname\relax
  \def\bibnamefont#1{#1}\fi
\expandafter\ifx\csname bibfnamefont\endcsname\relax
  \def\bibfnamefont#1{#1}\fi
\expandafter\ifx\csname citenamefont\endcsname\relax
  \def\citenamefont#1{#1}\fi
\expandafter\ifx\csname url\endcsname\relax
  \def\url#1{\texttt{#1}}\fi
\expandafter\ifx\csname urlprefix\endcsname\relax\def\urlprefix{URL }\fi
\providecommand{\bibinfo}[2]{#2}
\providecommand{\eprint}[2][]{\url{#2}}

\bibitem[{\citenamefont{Iroshnikov}(1963)}]{Iroshnikov63}
\bibinfo{author}{\bibfnamefont{R.~S.} \bibnamefont{Iroshnikov}},
  \bibinfo{journal}{Astron. Zh.} \textbf{\bibinfo{volume}{\boldVol{40}}},
  \bibinfo{pages}{742} (\bibinfo{year}{1963}), \bibinfo{note}{[Sov.\ Astron
  \textbf{7}, 566--571 (1964)]}.

\bibitem[{\citenamefont{Kraichnan}(1965)}]{Kraichnan65}
\bibinfo{author}{\bibfnamefont{R.~H.} \bibnamefont{Kraichnan}},
  \bibinfo{journal}{Phys.\ Fluids} \textbf{\bibinfo{volume}{\boldVol{8}}},
  \bibinfo{pages}{1385} (\bibinfo{year}{1965}).

\bibitem[{\citenamefont{Biskamp}(2003)}]{Biskamp-turb}
\bibinfo{author}{\bibfnamefont{D.}~\bibnamefont{Biskamp}},
  \emph{\bibinfo{title}{Magnetohydrodynamic Turbulence}}
  (\bibinfo{publisher}{CUP}, \bibinfo{address}{Cambridge},
  \bibinfo{year}{2003}).

\bibitem[{\citenamefont{Schekochihin et~al.}(2007)\citenamefont{Schekochihin,
  Cowley, and Dorland}}]{SchekochihinEA07-ppcf}
\bibinfo{author}{\bibfnamefont{A.~A.} \bibnamefont{Schekochihin}},
  \bibinfo{author}{\bibfnamefont{S.~C.} \bibnamefont{Cowley}},
  \bibnamefont{and} \bibinfo{author}{\bibfnamefont{W.}~\bibnamefont{Dorland}},
  \bibinfo{journal}{Plasma Phys.\ Controlled Fusion}
  \textbf{\bibinfo{volume}{\boldVol{49}}}, \bibinfo{pages}{A195}
  (\bibinfo{year}{2007}).

\bibitem[{\citenamefont{Robinson and Rusbridge}(1971)}]{RobinsonRusbridge71}
\bibinfo{author}{\bibfnamefont{D.}~\bibnamefont{Robinson}} \bibnamefont{and}
  \bibinfo{author}{\bibfnamefont{M.}~\bibnamefont{Rusbridge}},
  \bibinfo{journal}{Phys.\ Fluids} \textbf{\bibinfo{volume}{\boldVol{14}}},
  \bibinfo{pages}{2499} (\bibinfo{year}{1971}).

\bibitem[{\citenamefont{Zweben et~al.}(1979)\citenamefont{Zweben, Menyuk, and
  Taylor}}]{ZwebenEA79}
\bibinfo{author}{\bibfnamefont{S.}~\bibnamefont{Zweben}},
  \bibinfo{author}{\bibfnamefont{C.}~\bibnamefont{Menyuk}}, \bibnamefont{and}
  \bibinfo{author}{\bibfnamefont{R.}~\bibnamefont{Taylor}},
  \bibinfo{journal}{Phys.\ Rev.\ Lett.}
  \textbf{\bibinfo{volume}{\boldVol{42}}}, \bibinfo{pages}{1270}
  (\bibinfo{year}{1979}).

\bibitem[{\citenamefont{Matthaeus et~al.}(1990)\citenamefont{Matthaeus,
  Goldstein, and Roberts}}]{MattEA90}
\bibinfo{author}{\bibfnamefont{W.~H.} \bibnamefont{Matthaeus}},
  \bibinfo{author}{\bibfnamefont{M.~L.} \bibnamefont{Goldstein}},
  \bibnamefont{and} \bibinfo{author}{\bibfnamefont{D.~A.}
  \bibnamefont{Roberts}}, \bibinfo{journal}{J.\ Geophys.\ Res.}
  \textbf{\bibinfo{volume}{\boldVol{95}}}, \bibinfo{pages}{20\,673}
  (\bibinfo{year}{1990}).

\bibitem[{\citenamefont{Osman and Horbury}(2007)}]{OsmanHorbury07}
\bibinfo{author}{\bibfnamefont{K.~T.} \bibnamefont{Osman}} \bibnamefont{and}
  \bibinfo{author}{\bibfnamefont{T.~S.} \bibnamefont{Horbury}},
  \bibinfo{journal}{Astrophys.\ J.} \textbf{\bibinfo{volume}{\boldVol{654}}},
  \bibinfo{pages}{L103} (\bibinfo{year}{2007}).

\bibitem[{\citenamefont{Carbone et~al.}(1995)\citenamefont{Carbone, Malara, and
  Veltri}}]{CarboneEA95-Rij}
\bibinfo{author}{\bibfnamefont{V.}~\bibnamefont{Carbone}},
  \bibinfo{author}{\bibfnamefont{F.}~\bibnamefont{Malara}}, \bibnamefont{and}
  \bibinfo{author}{\bibfnamefont{P.}~\bibnamefont{Veltri}},
  \bibinfo{journal}{J.\ Geophys.\ Res.}
  \textbf{\bibinfo{volume}{\boldVol{100}}}, \bibinfo{pages}{1763}
  (\bibinfo{year}{1995}).

\bibitem[{\citenamefont{Dasso et~al.}(2005)\citenamefont{Dasso, Milano,
  Matthaeus, and Smith}}]{DassoEA05}
\bibinfo{author}{\bibfnamefont{S.}~\bibnamefont{Dasso}},
  \bibinfo{author}{\bibfnamefont{L.~J.} \bibnamefont{Milano}},
  \bibinfo{author}{\bibfnamefont{W.~H.} \bibnamefont{Matthaeus}},
  \bibnamefont{and} \bibinfo{author}{\bibfnamefont{C.~W.} \bibnamefont{Smith}},
  \bibinfo{journal}{Astrophys.\ J.} \textbf{\bibinfo{volume}{\boldVol{635}}},
  \bibinfo{pages}{L181} (\bibinfo{year}{2005}).

\bibitem[{\citenamefont{Podesta et~al.}(2007)\citenamefont{Podesta, Roberts,
  and Goldstein}}]{PodestaEA07-slopes}
\bibinfo{author}{\bibfnamefont{J.~J.} \bibnamefont{Podesta}},
  \bibinfo{author}{\bibfnamefont{D.~A.} \bibnamefont{Roberts}},
  \bibnamefont{and} \bibinfo{author}{\bibfnamefont{M.~L.}
  \bibnamefont{Goldstein}}, \bibinfo{journal}{Astrophys.\ J.}
  \textbf{\bibinfo{volume}{\boldVol{664}}}, \bibinfo{pages}{543}
  (\bibinfo{year}{2007}).

\bibitem[{\citenamefont{Bruno and Carbone}(2005)}]{BrunoCarbone05}
\bibinfo{author}{\bibfnamefont{R.}~\bibnamefont{Bruno}} \bibnamefont{and}
  \bibinfo{author}{\bibfnamefont{V.}~\bibnamefont{Carbone}},
  \bibinfo{journal}{Living Rev.\ Solar Phys.}
  \textbf{\bibinfo{volume}{\boldVol{2}}}, \bibinfo{pages}{{U}{R}{L}:
  http://www.livingreviews.org/lrsp} (\bibinfo{year}{2005}).

\bibitem[{\citenamefont{Cho et~al.}(2002)\citenamefont{Cho, Lazarian, and
  Vishniac}}]{ChoEA02-strongB0}
\bibinfo{author}{\bibfnamefont{J.}~\bibnamefont{Cho}},
  \bibinfo{author}{\bibfnamefont{A.}~\bibnamefont{Lazarian}}, \bibnamefont{and}
  \bibinfo{author}{\bibfnamefont{E.~T.} \bibnamefont{Vishniac}},
  \bibinfo{journal}{Astrophys.\ J.} \textbf{\bibinfo{volume}{\boldVol{564}}},
  \bibinfo{pages}{291} (\bibinfo{year}{2002}).

\bibitem[{\citenamefont{Goldreich and Sridhar}(1995)}]{GoldreichSridhar95}
\bibinfo{author}{\bibfnamefont{P.}~\bibnamefont{Goldreich}} \bibnamefont{and}
  \bibinfo{author}{\bibfnamefont{S.}~\bibnamefont{Sridhar}},
  \bibinfo{journal}{Astrophys.\ J.} \textbf{\bibinfo{volume}{\boldVol{438}}},
  \bibinfo{pages}{763} (\bibinfo{year}{1995}).

\bibitem[{\citenamefont{Boldyrev}(2006)}]{Boldyrev06}
\bibinfo{author}{\bibfnamefont{S.}~\bibnamefont{Boldyrev}},
  \bibinfo{journal}{Phys.\ Rev.\ Lett.}
  \textbf{\bibinfo{volume}{\boldVol{96}}}, \bibinfo{eid}{115002}
  (\bibinfo{year}{2006}),
  \urlprefix\url{http://link.aps.org/abstract/PRL/v96/e115002}.

\bibitem[{\citenamefont{Mason et~al.}(2008)\citenamefont{Mason, Cattaneo, and
  Boldyrev}}]{MasonEA08}
\bibinfo{author}{\bibfnamefont{J.}~\bibnamefont{Mason}},
  \bibinfo{author}{\bibfnamefont{F.}~\bibnamefont{Cattaneo}}, \bibnamefont{and}
  \bibinfo{author}{\bibfnamefont{S.}~\bibnamefont{Boldyrev}},
  \bibinfo{journal}{Phys.\ Rev.\ E} \textbf{\bibinfo{volume}{\boldVol{77}}},
  \bibinfo{eid}{036403} (\bibinfo{year}{2008}),
  \urlprefix\url{http://link.aps.org/abstract/PRE/v77/e036403}.

\bibitem[{\citenamefont{Higdon}(1984)}]{Higdon84}
\bibinfo{author}{\bibfnamefont{J.~C.} \bibnamefont{Higdon}},
  \bibinfo{journal}{Astrophys.\ J.} \textbf{\bibinfo{volume}{\boldVol{285}}},
  \bibinfo{pages}{109} (\bibinfo{year}{1984}).

\bibitem[{\citenamefont{Montgomery}(1982)}]{Mont82-strauss}
\bibinfo{author}{\bibfnamefont{D.~C.} \bibnamefont{Montgomery}},
  \bibinfo{journal}{Physica Scripta} \textbf{\bibinfo{volume}{\boldVol{T2/1}}},
  \bibinfo{pages}{83} (\bibinfo{year}{1982}).

\bibitem[{\citenamefont{Tu and Marsch}(1995)}]{TuMarsch95}
\bibinfo{author}{\bibfnamefont{C.-Y.} \bibnamefont{Tu}} \bibnamefont{and}
  \bibinfo{author}{\bibfnamefont{E.}~\bibnamefont{Marsch}},
  \bibinfo{journal}{Space Sci.\ Rev.} \textbf{\bibinfo{volume}{\boldVol{73}}},
  \bibinfo{pages}{1} (\bibinfo{year}{1995}).

\bibitem[{\citenamefont{Matthaeus et~al.}(1995)\citenamefont{Matthaeus, Bieber,
  and Zank}}]{MattEA95-unquiet}
\bibinfo{author}{\bibfnamefont{W.~H.} \bibnamefont{Matthaeus}},
  \bibinfo{author}{\bibfnamefont{J.~W.} \bibnamefont{Bieber}},
  \bibnamefont{and} \bibinfo{author}{\bibfnamefont{G.~P.} \bibnamefont{Zank}},
  \bibinfo{journal}{Rev.~Geophys.~Supp.}
  \textbf{\bibinfo{volume}{\boldVol{33}}}, \bibinfo{pages}{609}
  (\bibinfo{year}{1995}).

\bibitem[{\citenamefont{Horbury et~al.}(2005)\citenamefont{Horbury, Forman, and
  Oughton}}]{HorburyEA05}
\bibinfo{author}{\bibfnamefont{T.}~\bibnamefont{Horbury}},
  \bibinfo{author}{\bibfnamefont{M.~A.} \bibnamefont{Forman}},
  \bibnamefont{and} \bibinfo{author}{\bibfnamefont{S.}~\bibnamefont{Oughton}},
  \bibinfo{journal}{Plasma Phys.\ Controlled Fusion}
  \textbf{\bibinfo{volume}{\boldVol{47}}}, \bibinfo{pages}{B703}
  (\bibinfo{year}{2005}).

\bibitem[{\citenamefont{Taylor}(1938)}]{Taylor38}
\bibinfo{author}{\bibfnamefont{G.~I.} \bibnamefont{Taylor}},
  \bibinfo{journal}{Proc.\ Roy.\ Soc.\ Lond.\ A}
  \textbf{\bibinfo{volume}{\boldVol{164}}}, \bibinfo{pages}{476}
  (\bibinfo{year}{1938}).

\bibitem[{\citenamefont{Fredricks and Coroniti}(1976)}]{FredricksCoroniti76}
\bibinfo{author}{\bibfnamefont{R.~W.} \bibnamefont{Fredricks}}
  \bibnamefont{and} \bibinfo{author}{\bibfnamefont{F.~V.}
  \bibnamefont{Coroniti}}, \bibinfo{journal}{J.\ Geophys.\ Res.}
  \textbf{\bibinfo{volume}{\boldVol{81}}}, \bibinfo{pages}{5591}
  (\bibinfo{year}{1976}).

\bibitem[{\citenamefont{Bieber et~al.}(1996)\citenamefont{Bieber, Wanner, and
  Matthaeus}}]{BieberEA96}
\bibinfo{author}{\bibfnamefont{J.~W.} \bibnamefont{Bieber}},
  \bibinfo{author}{\bibfnamefont{W.}~\bibnamefont{Wanner}}, \bibnamefont{and}
  \bibinfo{author}{\bibfnamefont{W.~H.} \bibnamefont{Matthaeus}},
  \bibinfo{journal}{J.\ Geophys.\ Res.}
  \textbf{\bibinfo{volume}{\boldVol{101}}}, \bibinfo{pages}{2511}
  (\bibinfo{year}{1996}).

\bibitem[{\citenamefont{Balogh et~al.}(1999)\citenamefont{Balogh, Forsyth,
  Lucek, Horbury, and Smith}}]{BaloghEA99}
\bibinfo{author}{\bibfnamefont{A.}~\bibnamefont{Balogh}},
  \bibinfo{author}{\bibfnamefont{R.~J.} \bibnamefont{Forsyth}},
  \bibinfo{author}{\bibfnamefont{E.~A.} \bibnamefont{Lucek}},
  \bibinfo{author}{\bibfnamefont{T.~S.} \bibnamefont{Horbury}},
  \bibnamefont{and} \bibinfo{author}{\bibfnamefont{E.~J.} \bibnamefont{Smith}},
  \bibinfo{journal}{Geophys.\ Res.\ Lett.}
  \textbf{\bibinfo{volume}{\boldVol{26}}}, \bibinfo{pages}{631}
  (\bibinfo{year}{1999}).

\bibitem[{\citenamefont{Mc{C}omas et~al.}(2000)\citenamefont{Mc{C}omas,
  Barraclough, Funsten, Gosling, Santiago-{Mu\~noz}, Skoug, Goldstein,
  Neugebauer, Riley, and Balogh}}]{McComasEA00}
\bibinfo{author}{\bibfnamefont{D.~J.} \bibnamefont{Mc{C}omas}},
  \bibinfo{author}{\bibfnamefont{L.}~\bibnamefont{Barraclough}},
  \bibinfo{author}{\bibfnamefont{H.~O.} \bibnamefont{Funsten}},
  \bibinfo{author}{\bibfnamefont{J.~T.} \bibnamefont{Gosling}},
  \bibinfo{author}{\bibfnamefont{E.}~\bibnamefont{Santiago-{Mu\~noz}}},
  \bibinfo{author}{\bibfnamefont{R.~M.} \bibnamefont{Skoug}},
  \bibinfo{author}{\bibfnamefont{B.~E.} \bibnamefont{Goldstein}},
  \bibinfo{author}{\bibfnamefont{M.}~\bibnamefont{Neugebauer}},
  \bibinfo{author}{\bibfnamefont{P.}~\bibnamefont{Riley}}, \bibnamefont{and}
  \bibinfo{author}{\bibfnamefont{A.}~\bibnamefont{Balogh}},
  \bibinfo{journal}{J.\ Geophys.\ Res.}
  \textbf{\bibinfo{volume}{\boldVol{105}}}, \bibinfo{pages}{10\,419}
  (\bibinfo{year}{2000}).

\bibitem[{\citenamefont{Horbury and Balogh}(2001)}]{HorburyBalogh01}
\bibinfo{author}{\bibfnamefont{T.}~\bibnamefont{Horbury}} \bibnamefont{and}
  \bibinfo{author}{\bibfnamefont{A.}~\bibnamefont{Balogh}},
  \bibinfo{journal}{J.\ Geophys.\ Res.}
  \textbf{\bibinfo{volume}{\boldVol{106}}}, \bibinfo{pages}{15\,929}
  (\bibinfo{year}{2001}).

\bibitem[{\citenamefont{Smith et~al.}(1995)\citenamefont{Smith, Balogh,
  Neugebauer, and Mc{C}omas}}]{SmithEA95d}
\bibinfo{author}{\bibfnamefont{E.~J.} \bibnamefont{Smith}},
  \bibinfo{author}{\bibfnamefont{A.}~\bibnamefont{Balogh}},
  \bibinfo{author}{\bibfnamefont{M.}~\bibnamefont{Neugebauer}},
  \bibnamefont{and}
  \bibinfo{author}{\bibfnamefont{D.}~\bibnamefont{Mc{C}omas}},
  \bibinfo{journal}{Geophys.\ Res.\ Lett.}
  \textbf{\bibinfo{volume}{\boldVol{22}}}, \bibinfo{pages}{3381}
  (\bibinfo{year}{1995}).

\bibitem[{\citenamefont{Horbury et~al.}(1995)\citenamefont{Horbury, Balogh,
  Forsyth, and Smith}}]{HorburyEA95c}
\bibinfo{author}{\bibfnamefont{T.}~\bibnamefont{Horbury}},
  \bibinfo{author}{\bibfnamefont{A.}~\bibnamefont{Balogh}},
  \bibinfo{author}{\bibfnamefont{R.~J.} \bibnamefont{Forsyth}},
  \bibnamefont{and} \bibinfo{author}{\bibfnamefont{E.~J.} \bibnamefont{Smith}},
  \bibinfo{journal}{Geophys.\ Res.\ Lett.}
  \textbf{\bibinfo{volume}{\boldVol{22}}}, \bibinfo{pages}{3405}
  (\bibinfo{year}{1995}).

\bibitem[{\citenamefont{Percival and Walden}(2000)}]{PercivalWalden}
\bibinfo{author}{\bibfnamefont{D.~B.} \bibnamefont{Percival}} \bibnamefont{and}
  \bibinfo{author}{\bibfnamefont{A.~T.} \bibnamefont{Walden}},
  \emph{\bibinfo{title}{Wavelet Methods for Time Series Analysis}}
  (\bibinfo{publisher}{CUP}, \bibinfo{address}{Cambridge},
  \bibinfo{year}{2000}).

\bibitem[{\citenamefont{Farge}(1992)}]{Farge92}
\bibinfo{author}{\bibfnamefont{M.}~\bibnamefont{Farge}},
  \bibinfo{journal}{Ann.\ Rev.\ Fluid Mech.}
  \textbf{\bibinfo{volume}{\boldVol{24}}}, \bibinfo{pages}{395}
  (\bibinfo{year}{1992}).

\bibitem[{\citenamefont{Torrence and Compo}(1998)}]{TorrenceCompo98}
\bibinfo{author}{\bibfnamefont{C.}~\bibnamefont{Torrence}} \bibnamefont{and}
  \bibinfo{author}{\bibfnamefont{G.~P.} \bibnamefont{Compo}},
  \bibinfo{journal}{Bull.\ Am.\ Meteorol.\ Soc.}
  \textbf{\bibinfo{volume}{\boldVol{79}}}, \bibinfo{pages}{61}
  (\bibinfo{year}{1998}).

\bibitem[{\citenamefont{Alexandrova et~al.}(2006)\citenamefont{Alexandrova,
  Mangeney, Maksimovic, Cornilleau-{Wehrlin}, Bosques, and
  Andr\'e}}]{AlexandrovaEA06}
\bibinfo{author}{\bibfnamefont{O.}~\bibnamefont{Alexandrova}},
  \bibinfo{author}{\bibfnamefont{A.}~\bibnamefont{Mangeney}},
  \bibinfo{author}{\bibfnamefont{M.}~\bibnamefont{Maksimovic}},
  \bibinfo{author}{\bibfnamefont{N.}~\bibnamefont{Cornilleau-{Wehrlin}}},
  \bibinfo{author}{\bibfnamefont{J.-M.} \bibnamefont{Bosques}},
  \bibnamefont{and} \bibinfo{author}{\bibfnamefont{M.}~\bibnamefont{Andr\'e}},
  \bibinfo{journal}{J.\ Geophys.\ Res.}
  \textbf{\bibinfo{volume}{\boldVol{111}}}, \bibinfo{eid}{A12208}
  (\bibinfo{year}{2006}).

\bibitem[{\citenamefont{Smith}(2003)}]{Smith-sw10}
\bibinfo{author}{\bibfnamefont{C.~W.} \bibnamefont{Smith}}, in
  \emph{\bibinfo{booktitle}{Solar Wind Ten}}, edited by
  \bibinfo{editor}{\bibfnamefont{M.}~\bibnamefont{Velli}},
  \bibinfo{editor}{\bibfnamefont{R.}~\bibnamefont{Bruno}}, \bibnamefont{and}
  \bibinfo{editor}{\bibfnamefont{F.}~\bibnamefont{Malara}}
  (\bibinfo{publisher}{AIP}, \bibinfo{address}{Melville, NY},
  \bibinfo{year}{2003}), vol. \bibinfo{volume}{679}, pp.
  \bibinfo{pages}{413--416}.

\bibitem[{\citenamefont{Leamon et~al.}(1998)\citenamefont{Leamon,
  \SortNoop{aaa}Smith, Ness, Matthaeus, and Wong}}]{LeamonEA98a}
\bibinfo{author}{\bibfnamefont{R.~J.} \bibnamefont{Leamon}},
  \bibinfo{author}{\bibfnamefont{C.~W.} \bibnamefont{\SortNoop{aaa}Smith}},
  \bibinfo{author}{\bibfnamefont{N.~F.} \bibnamefont{Ness}},
  \bibinfo{author}{\bibfnamefont{W.~H.} \bibnamefont{Matthaeus}},
  \bibnamefont{and} \bibinfo{author}{\bibfnamefont{H.~K.} \bibnamefont{Wong}},
  \bibinfo{journal}{J.\ Geophys.\ Res.}
  \textbf{\bibinfo{volume}{\boldVol{103}}}, \bibinfo{pages}{4775}
  (\bibinfo{year}{1998}).

\bibitem[{\citenamefont{Boldyrev}(2005)}]{Boldyrev05}
\bibinfo{author}{\bibfnamefont{S.}~\bibnamefont{Boldyrev}},
  \bibinfo{journal}{Astrophys.\ J.} \textbf{\bibinfo{volume}{\boldVol{626}}},
  \bibinfo{pages}{L37} (\bibinfo{year}{2005}).

\bibitem[{\citenamefont{Narita et~al.}(2007)\citenamefont{Narita, Glassmeier,
  Goldstein, and Treumann}}]{NaritaEA07-riverside}
\bibinfo{author}{\bibfnamefont{Y.}~\bibnamefont{Narita}},
  \bibinfo{author}{\bibfnamefont{K.-H.} \bibnamefont{Glassmeier}},
  \bibinfo{author}{\bibfnamefont{M.~L.} \bibnamefont{Goldstein}},
  \bibnamefont{and} \bibinfo{author}{\bibfnamefont{R.~A.}
  \bibnamefont{Treumann}}, in \emph{\bibinfo{booktitle}{Turbulence and
  Nonlinear Processes in Astrophysical Plasmas}}, edited by
  \bibinfo{editor}{\bibfnamefont{D.}~\bibnamefont{Shaikh}} \bibnamefont{and}
  \bibinfo{editor}{\bibfnamefont{G.~P.} \bibnamefont{Zank}}
  (\bibinfo{publisher}{AIP}, \bibinfo{year}{2007}), vol.
  \bibinfo{volume}{CP932}, pp. \bibinfo{pages}{215--220}, \bibinfo{note}{6th
  Annual International Astrophysical Conference}.

\bibitem[{\citenamefont{Sahraoui et~al.}(2006)\citenamefont{Sahraoui, Belmont,
  Rezeau, Cornilleau-{Wehrlin}, Pin\c{c}on, and Balogh}}]{SahraouiEA06}
\bibinfo{author}{\bibfnamefont{F.}~\bibnamefont{Sahraoui}},
  \bibinfo{author}{\bibfnamefont{G.}~\bibnamefont{Belmont}},
  \bibinfo{author}{\bibfnamefont{L.}~\bibnamefont{Rezeau}},
  \bibinfo{author}{\bibfnamefont{N.}~\bibnamefont{Cornilleau-{Wehrlin}}},
  \bibinfo{author}{\bibfnamefont{J.~L.} \bibnamefont{Pin\c{c}on}},
  \bibnamefont{and} \bibinfo{author}{\bibfnamefont{A.}~\bibnamefont{Balogh}},
  \bibinfo{journal}{Phys.\ Rev.\ Lett.}
  \textbf{\bibinfo{volume}{\boldVol{96}}}, \bibinfo{eid}{075002}
  (\bibinfo{year}{2006}),
  \urlprefix\url{http://link.aps.org/abstract/PRL/v96/e075002}.

\bibitem[{\citenamefont{M\"uller and Grappin}(2005)}]{MullerGrappin05}
\bibinfo{author}{\bibfnamefont{W.-C.} \bibnamefont{M\"uller}} \bibnamefont{and}
  \bibinfo{author}{\bibfnamefont{R.}~\bibnamefont{Grappin}},
  \bibinfo{journal}{Phys.\ Rev.\ Lett.}
  \textbf{\bibinfo{volume}{\boldVol{95}}}, \bibinfo{eid}{114502}
  (\bibinfo{year}{2005}),
  \urlprefix\url{http://link.aps.org/abstract/PRL/v95/e114502}.

\bibitem[{\citenamefont{Cho and Vishniac}(2000)}]{ChoVishniac00-aniso}
\bibinfo{author}{\bibfnamefont{J.}~\bibnamefont{Cho}} \bibnamefont{and}
  \bibinfo{author}{\bibfnamefont{E.~T.} \bibnamefont{Vishniac}},
  \bibinfo{journal}{Astrophys.\ J.} \textbf{\bibinfo{volume}{\boldVol{539}}},
  \bibinfo{pages}{273} (\bibinfo{year}{2000}).

\bibitem[{\citenamefont{Maron and Goldreich}(2001)}]{MaronGoldreich01}
\bibinfo{author}{\bibfnamefont{J.}~\bibnamefont{Maron}} \bibnamefont{and}
  \bibinfo{author}{\bibfnamefont{P.}~\bibnamefont{Goldreich}},
  \bibinfo{journal}{Astrophys.\ J.} \textbf{\bibinfo{volume}{\boldVol{554}}},
  \bibinfo{pages}{1175} (\bibinfo{year}{2001}).

\end{thebibliography}
 \newcommand{\boldVol}[1] {\textbf{#1}} 
  \newcommand{\SortNoop}[1] {} 
  \newcommand{\au} {{A}{U}\ } 
  \newcommand{\AU} {{A}{U}\ } 
  \newcommand{\MHD} {{M}{H}{D}\ } 
  \newcommand{\mhd} {{M}{H}{D}\ } 
  \newcommand{\RMHD} {{R}{M}{H}{D}\ } 
  \newcommand{\rmhd} {{R}{M}{H}{D}\ } 
  \newcommand{\wkb} {{W}{K}{B}\ } 
  \newcommand{\alfven} {{A}lfv\'en\ } 
  \newcommand{\Alfven} {{A}lfv\'en\ } 
  \newcommand{\alfvenic} {{A}lfv\'enic\ } 
  \newcommand{\Alfvenic} {{A}lfv\'enic\ }

\end{document}